\documentclass[fleqn, twocolumn, 11pt]{wlscirep}
\usepackage[utf8]{inputenc}
\usepackage[T1]{fontenc}
\title{Bridging Grain Mapping and Dark Field X-ray Microscopy for Multiscale Diffraction Imaging}
\usepackage{gensymb}
\usepackage{multirow}
\usepackage{ragged2e}
\def\tablename{Table}%
\usepackage{float}
\usepackage{amsmath} 
\usepackage{epstopdf}           %
\usepackage{lmodern}
\usepackage{siunitx}

\usepackage{lineno}
\usepackage[figuresright]{rotating}
\usepackage{color,soul}

\usepackage{hyperref}
\hypersetup{
    colorlinks=true,
    linkcolor=blue,
    filecolor=magenta,      
    urlcolor=cyan,
    pdftitle={Overleaf Example},
    pdfpagemode=FullScreen,
    }
    
\usepackage{subfigure}

\author[1]{A. Shukla}
\author[1,*]{C. Yildirim}
\author[1]{J. A. D. Ball}
\author[1]{C. Detlefs}
\author[2]{A. A. W. Cretton}
\author[1]{M. Sarkis}
\author[2]{M. La Bella}
\author[1,4]{W. Ludwig}
\author[3]{Y. Zhang} 
\author[3]{N. A. Henningsson}

\affil[1]{European Synchrotron Radiation Facility, 71 Avenue des Martyrs, CS40220, 38043 Grenoble Cedex 9, France}
\affil[2]{Department of Physics, Technical University of Denmark, Fysikvej, 2800 Kgs. Lyngby, Denmark}
\affil[3]{Department of Civil and Mechanical Engineering, Technical University of Denmark, 2800 Kgs. Lyngby, Denmark}
\affil[4]{Univ Lyon,CNRS, INSA Lyon, Université Claude Bernard Lyon 1, MATEIS, UMR5510, 69621 Villeurbanne, France}

\affil[*]{can.yildirim@esrf.fr}
\date{\today}

\begin{abstract}

Resolving how defects emerge and interact within the hierarchical structure of polycrystalline materials remains a core challenge in materials science. Grain-mapping methods such as three-dimensional X-ray diffraction (3DXRD) and diffraction contrast tomography (DCT) provide essential mesoscale context but lack the resolution to image lattice defects. Conversely, high-resolution methods like Dark Field X-ray Microscopy (DFXM) capture lattice distortions but not the surrounding microstructure. Here, we introduce a transferable framework that unifies these complementary approaches into a single, non-destructive workflow. Enabled by open-source software, the method translates grain orientation and position data into precise goniometer settings for DFXM imaging without dismounting or reorienting the sample. Applied to an iron polycrystal containing $\approx$ 1100 grains, DFXM motor positions were calculated for all grains within seconds, enabling on-the-fly targeting of specific grains. This allows reproducible zooming from the millimetre-scale aggregate to individual dislocations. We resolve three-dimensional misorientation fields across grain boundaries with 36 nm pixel size, directly capturing grain–grain interactions within their microstructural context. Finally, we show transferability from LabDCT to synchrotron and XFEL platforms, enabling new ways of studying defect interactions across scales.

\end{abstract}

\begin{document}
\flushbottom
\maketitle



Understanding how defects evolve and affect macroscopic properties is a central challenge in materials science. Most structural and functional materials are polycrystalline, with grains bounded by interfaces and filled with subgrains and dislocations. This hierarchical microstructure spans from atomic to millimeter scales and controls key behaviors like strength, toughness, and conductivity. Bridging these scales in 3D, while capturing both context and defect-level detail, demands a new generation of non-destructive, high-resolution techniques.

Conventional microscopy approaches such as transmission electron microscopy (TEM) and electron backscatter diffraction (EBSD) have advanced our understanding of the link between crystallographic orientation and defect structures, but they are surface-sensitive or require destructive sample preparation. As such, they are not well suited to in-situ bulk investigations. In contrast, synchrotron X-ray-based grain mapping techniques, including three-dimensional X-ray diffraction microscopy (3DXRD) \cite{poulsen2004three, Schmidt229, hefferan2012}, diffraction contrast tomography (DCT) \cite{king2008observations, sun2018}, and differential aperture X-ray microscopy (DAXM) \cite{larson_three-dimensional_2002,xu2017direct, knipschildt2025}, offer 3D, non-destructive mapping of grain structure, enabling study of thousands of grains, embedded deep within polycrystalline samples.

Each method, however, involves trade-offs. For example, 3DXRD and DCT provide fast 3D orientation mapping of a polycrystal but are limited in spatial resolution (\( 1\,\mu\mathrm{m} \)) and have reduced applicability for  highly  strained materials. Scanning-based variants such as scanning 3DXRD \cite{henningsson2024, shukla2024grain,Hayashi:ks5460,Henningsson:nb5257, Henningsson:nb5298} or texture tomography \cite{frewein2024texture, carlsen2024x} improve spatial resolution but are time intensive. High-resolution approaches like Bragg ptychography \cite{hruszkewycz2017high} and Bragg coherent diffraction imaging \cite{richard2022bragg} reach the nanometer regime, but are limited to undeformed grains with low dislocation density.

Dark Field X-ray Microscopy (DFXM) has emerged as a powerful alternative by combining high spatial resolution ($\sim$100\,nm) and sub-microradian angular resolution with bulk imaging capabilities \cite{Simons2015, yildirim2020probing, Poulsen2017, isern2025}. It enables full-field, real-space mapping of strain and orientation variations within individual grains, revealing dislocation networks, strain gradients, and cellular structure \cite{zelenika20243d, zelenika202, extensive, cretton2025}. Virtual 2D slicing using a line-focused beam allows rapid mapping in the subsecond regime, as demonstrated in time-resolved experiments on dislocation dynamics in aluminum near its melting point \cite{dresselhaus2021situ}. Today, a full 3D orientation and strain mapping of a single grain now typically takes about an hour or more, depending on its size and angular spread. An alternative for 3D mapping is magnified topotomography (MTT) \cite{Simons2015,Ludwig2001}, where a full-field box-shaped beam illuminates the grain while it rotates about the diffraction vector. Recent advances using pink beam DFXM \cite{yildirim2025pink} have increased acquisition speed more than twentyfold. Yet, DFXM has a critical limitation: it provides high-resolution imaging only for one Bragg reflection of a single grain at a time, with no information about the surrounding microstructure. Without prior information about grain orientation and position,  the targeted selection of specific reflections associated with grains of interest is not feasible, which prevents the reconstruction of complete strain tensors using DFXM and grain orientaiton information , as demonstrated in previous works \cite{Detlefs:xx5078, HENNINGSSON2025106277}. 
Although the combination of DFXM with coarser-grained techniques such as DCT or 3DXRD has been demonstrated in a few exceptional cases \cite{gustafson2020, gustafson2023,doi:10.1126/science.adv3460}, these efforts were limited to specially prepared samples and involved complex, non-repeatable procedures. As a result, no unified or robust framework currently exists for correlating bulk polycrystalline microstructures with high-resolution imaging of lattice defects, dislocations, and subgrains embedded within each grain of the polycrystalline aggregate.
The field is in urgent need of a seamless, repeatable multiscale workflow that enables targeted, high-resolution imaging based on grain-resolved pre-characterization, without removing or repositioning the sample. 

Here, we introduce a general framework that connects coarse microstructure characterization methods, 3DXRD, DCT, phase contrast tomography (PCT), and LabDCT \cite{10.1063/1.3100200,oh2025taking} , with DFXM in a single workflow. This allows zooming in from a polycrystalline aggregate to specific grains and revealing dislocation networks, subgrains, and elastic strains with high resolution. We demonstrate this framework, through measurements on a polycrystalline body-centered cubic (BCC) iron sample, mapping ~1100 grains using 3DXRD and DCT, and then zooming into a selected grain neighborhood (triple junction) with DFXM. This zoom out, zoom in approach enabled by our framework allows the direct observation of subtle orientation and strain variations \textit{within targeted individual grains}, with 100 nm spatial resolution.

At the core of this framework is our open-source software package, \texttt{crispy}, which computes precise goniometer motor positions in an X-ray diffractometer from grain orientation and position data for thousands of grains with sufficiently fast processing (within seconds) to enable experimental feedback. This automates the process of bringing selected grains into their respective Bragg condition for DFXM experiments with 
high precision, thereby facilitating targeted imaging of grain boundaries or texture components without dismounting or repositioning the sample.

This approach bridges the gap from millimetre-scale polycrystalline aggregates to nanoscale dislocation structures, allowing spatially targeted, high-resolution imaging across lab, synchrotron, and X-ray Free-Electron Laser (XFEL) platforms. \cite{dresselhaus2023simultaneous}

\section*{\textbf{Results and Discussion}}

\subsection*{Overview of Multiscale Imaging: 3DXRD/DCT/PCT/DFXM}

\begin{figure*}[h!]
    \centering
    \includegraphics[width=0.7\linewidth]{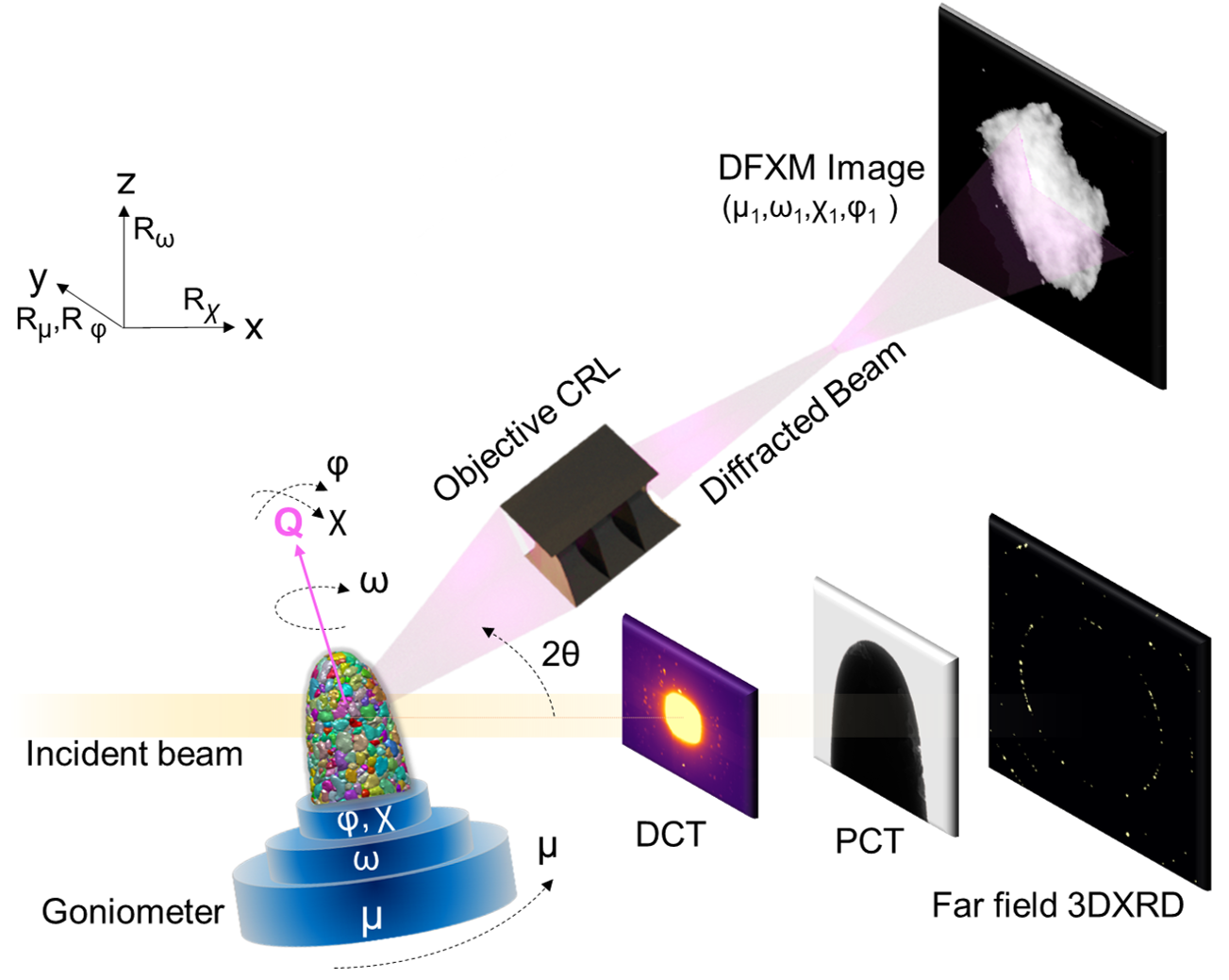}
    \caption{
        \justifying
        \textbf{Schematic of the multimodal and multiscale X-ray diffraction microscopy setup.} 
        \small A monochromatic X-ray beam illuminates a polycrystalline sample mounted on a high-precision goniometer equipped with four rotation stages: $\mu$, $\omega$, $\chi$, and $\varphi$. The goniometer enables full orientation control of the sample relative to the incident beam and the diffraction vector $\mathbf{Q}$. Four complementary measurement modalities are integrated without dismounting the sample. A near-field detector coupled with a scintillator and visible light optics is used to perform Diffraction Contrast Tomography (DCT), capturing grain morphology and orientation. This detector can also be used to measure Phase Contrast Tomography (PCT) when placed farther from the sample. A downstream 2D detector with a large field of view records far-field 3D X-ray diffraction (3DXRD) patterns for grain-resolved strain and orientation analysis. In the diffracted beam path, a high-resolution far-field detector positioned approximately 5 meters from the sample and aligned with the compound refractive lenses (CRLs) collects Dark Field X-ray Microscopy (DFXM) images from a grain of interest within the illuminated volume. This configuration allows multiscale characterization of crystalline microstructures, combining grain level morphology, orientation, and intra-grain defect contrast in a single experimental setup. Lab coordinate axes are indicated for clarity.
    }
    \label{fig:schematics}
\end{figure*}

Our multiscale imaging workflow begins with a coarse grain mapping step using either 3DXRD, DCT or LabDCT as a pre-characterization step. We demonstrate this workflow implementated at the ID03 beamline of European Synchrotron Radiation Facility (ESRF). In a far-field 3DXRD experiment, far-field diffraction patterns are recorded on an area detector placed downstream of the sample, providing rapid access to grain-averaged orientation, strain, and center of mass positions. Alternatively, a near-field detector coupled with a scintillator and optical system can be used for a DCT experiment, offering more accurate grain shapes at the cost of longer acquisition and data processing times. For completeness, both modalities were employed in this study. The diffraction data can be complemented with PCT data at the , which can be collected on the near-field detector.

Once grains are indexed and associated with their orientation matrices using 3DXRD, DCT, or LabDCT, our open-source package \texttt{crispy} computes the required goniometer motor positions ($\mu$, $\omega$, $\chi$, $\varphi$) to bring any selected grain into the Bragg condition for DFXM. See Supplementary Information (SI) for details on coordinate transformations and goniometer motor position calculations. As shown in Fig.~\ref{fig:schematics}, the diffracted signal from a specific diffraction vector \(\boldsymbol{Q}\) of a grain of interest corresponding to a set of Miller indices (hkl) is directed through the objective optics and collected on the high-resolution far-field DFXM detector for the dark field experiment. Based on the specific application requirements, an appropriate mode of DFXM measurement can be selected from several available configurations, including but not limited to using monochromatic or pink-beam in projection mode, layer-scanning mode, or MTT mode (see Methods for DFXM operation modes). 

\subsection*{Coarser Scale Grain Mapping with 3DXRD \& DCT}

\begin{figure}[t]
    \centering
    \includegraphics[width=1\linewidth]{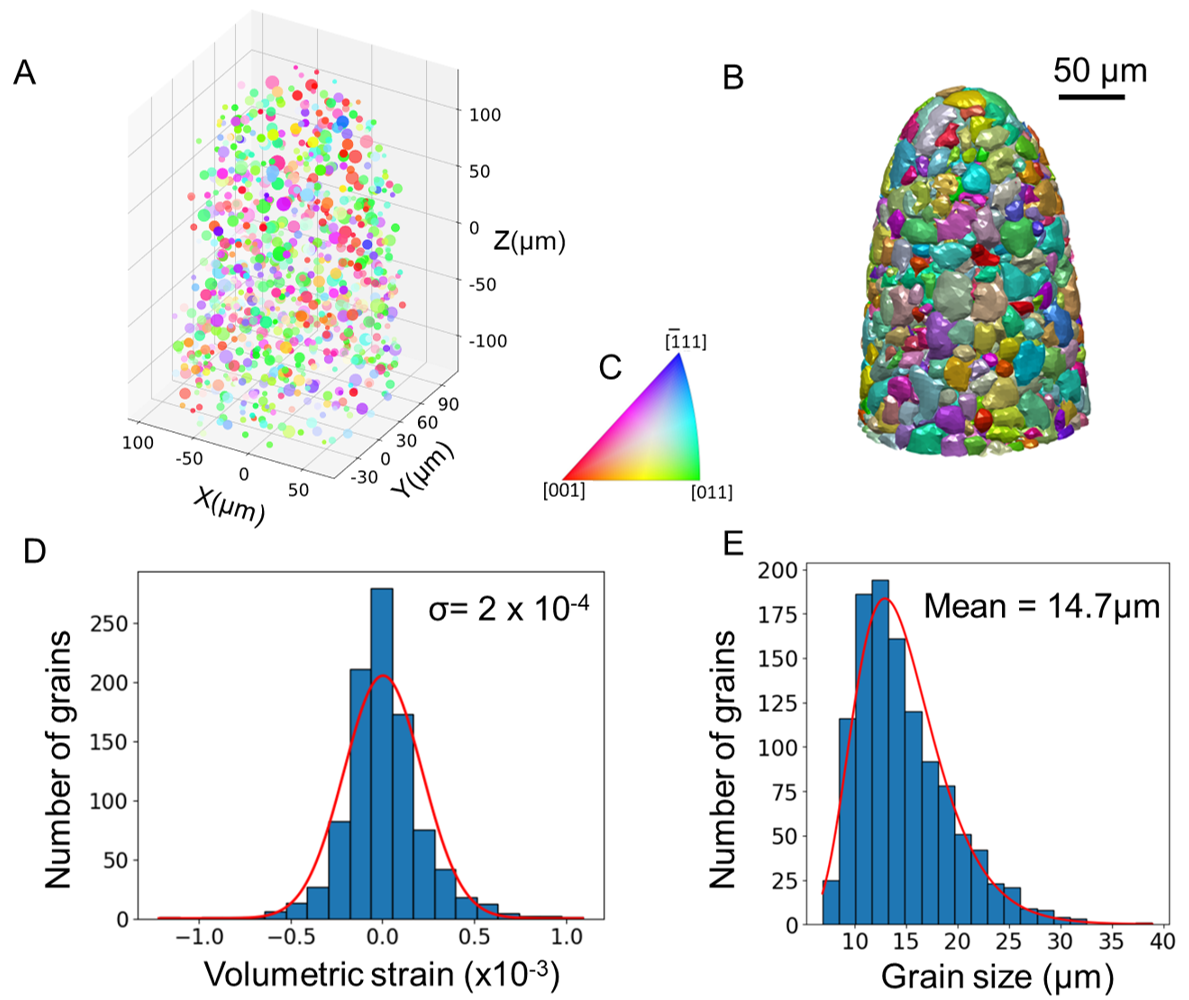}
    \caption{
    \textbf{Grain-resolved structural and statistical analysis using 3DXRD and DCT.} 
    \justifying
    \small (A) Reconstructed 3DXRD point cloud showing the center of mass positions and orientations of individual grains, colored by inverse pole figure (IPF-Z) according to the colormap in (C). The size of each point scales with the volume of the corresponding grain.   
    (B) Grain morphology obtained from DCT reconstruction.
    (D) Distribution of grain-averaged volumetric strain measured by 3DXRD. The red curve shows a Gaussian fit.   
    (E) Grain size distribution derived from the DCT reconstruction. The red curve shows a log-normal fit.  
    }
    \label{fig:3dxrd_dct}
\end{figure}

Fig.~\ref{fig:3dxrd_dct}A and B show the reconstructed microstructures of a fully recrystallized pure iron sample, obtained from 3DXRD and DCT measurements, respectively (see Methods for sample details). Both reconstructions reveal the conical geometry of the sample, which agrees with the shape observed via PCT (Figure S1A in the SI). DCT analysis predicted approximately 1100 grains within the measured volume. Of these, 949 were successfully indexed using 3DXRD while measuring two diffraction rings on the far-field detector. 


The sample exhibited observable texture as seen in the inverse pole figure (Figure S1B in SI (SI)). Figure \ref{fig:3dxrd_dct}D shows the distribution of grain-averaged volumetric strain (calculated as trace of the strain tensor) extracted from the 3DXRD data. A gaussian fit to the distribution (red curve) yields a standard deviation of $2 \times 10^{-4}$, indicating a narrow spread around zero strain, characteristic of the recrystallized state of the microstructure \cite{zhang2022local}. Figure \ref{fig:3dxrd_dct}E shows the grain size distribution derived from the DCT reconstruction. A log-normal distribution fit (red curve) suggests a mean grain diameter of \( 14.7\,\mu\mathrm{m} \) consistent with observations from studies on a similar sample \cite{zhang2022local}. 

\subsection*{Fast 2D Imaging of Neighboring Grains}

\begin{figure*}[h!]
    \centering
    \includegraphics[width=0.8\linewidth]{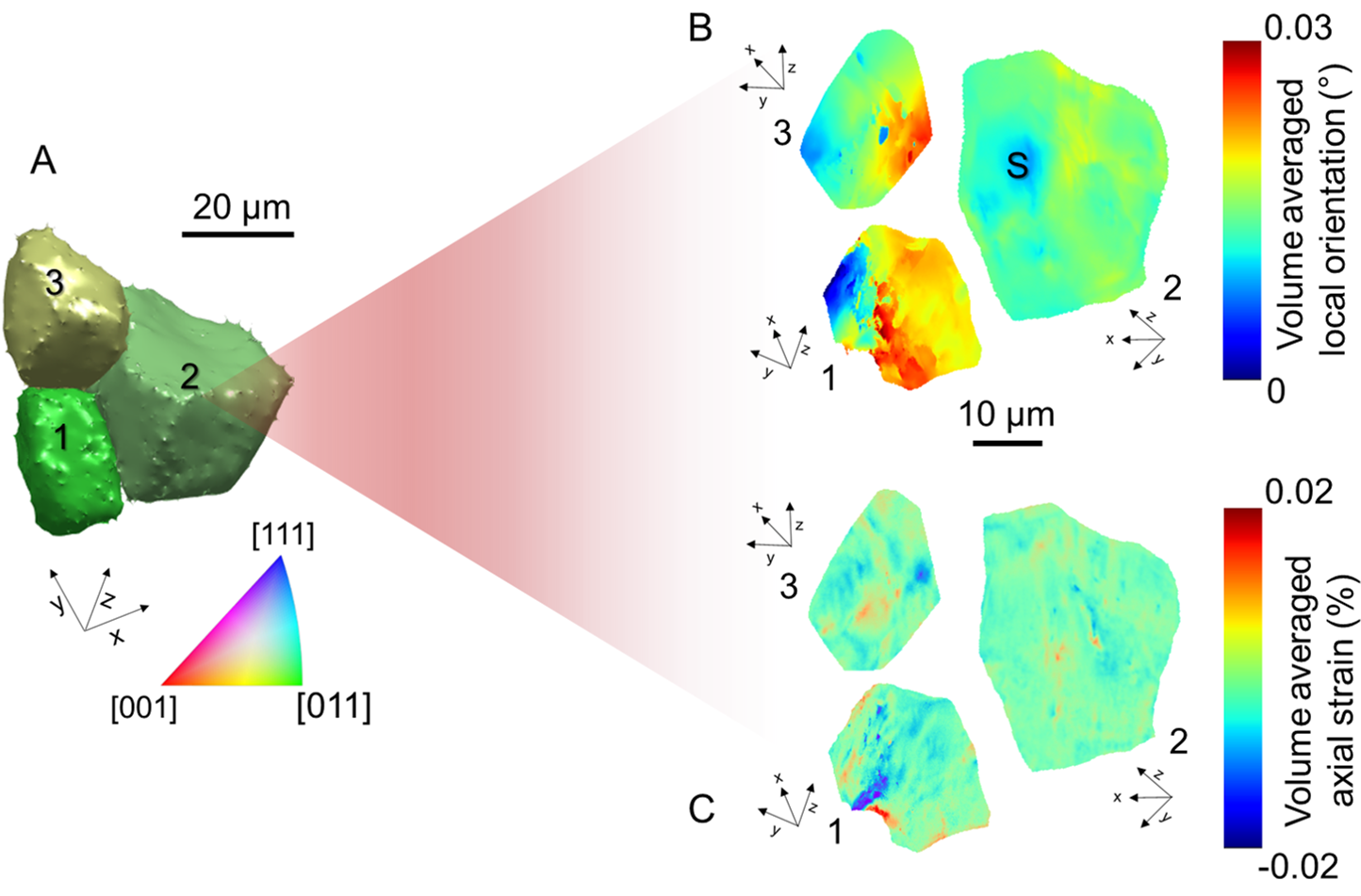}
    \caption{
    \justifying
    \textbf{Spatially resolved intragranular orientation and strain mapping of neighboring grains.}
    \small
    (A) Reconstructed grain orientation map from DCT, colored by IPF-Z. (B) Local orientation and (C) axial strain maps along the diffraction vector $\boldsymbol{Q}_{111}$, of a 2D projection of the grains, using a box beam. Grains 1-3 are labeled, with their respective rotations given in the laboratory coordinate system. 
    }
    \label{fig:projection_imaging}
\end{figure*}

Next, we zoom into a specific grain neighborhood in the polycrystalline aggregate using DFXM. Specifically, we chose a cluster of three neighboring grains near the center of the sample, forming a triple junction—a feature known to influence grain growth in recrystallized microstructures and deformation twinning \cite{zhang2020grain, lee2025}. The corresponding DCT grain map for the three grains is shown in Fig.\ref{fig:projection_imaging}A. After goniometer motor position alignment using 3DXRD data and the developed algorithm (See SI for implementation details), the three grains were imaged with DFXM. Here a full field box-beam 'projection mode' was used such that the DFXM detector images corresponded to 2D projections of integrated intensity from the 3D diffraction grain volumes. The local (volume averaged) intra-grain lattice orientation was calculated from these measurements and are shown in Fig. \ref{fig:projection_imaging}B. The error between the simulated goniometer angles from \texttt{crispy} and the experimentally determined motor positions is shown in Table 1 of the SI. All errors are within $0.1^\circ$, which matches the angular stepsize of the 3DXRD measurement and are on the order of misorientation spread within each grain. This close agreement confirms the reliability of the overall multiscale workflow and alignment framework.

DFXM local orientation maps (calculated using the center of mass map of the rocking angle \cite{darfix}) show the local misorientation within each grain relative to its nominal lattice plane normal while DFXM strain maps show the local variation in axial strain \cite{Simons2015}. In this example, the local orientation and axial strain data within all of the three neighboring grains was measured within 20 minutes using the DFXM projection mode. It is important to note that, in a DFXM experiment, each grain is measured within its own local coordinate system, as illustrated in Figure ~\ref{fig:projection_imaging}B. All the three neighboring grains exhibit distinct intragranular features and small orientation variations below 0.02$^\circ$ and subgrain boundaries that persist even after recrystallization,  in agreement with previous observations in aluminum \cite{Ahl2017}. 

The distinct misorientation pattern of Grain 1 is seen in Fig.~\ref{fig:projection_imaging}B with the left side of the grain showing higher local misorientation than the right. Similarly, Grain 3 exhibits increased misorientation near the grain boundary, while Grain 2 appears mostly uniform. A subgrain, denoted as S, is identified within Grain 2. This feature is better resolved in the weak-beam dark-field image, captured at the tails of grain's rocking curve \cite{Borgi:nb5396} (SI Fig. S2), revealing contrast between the defect-free matrix and the subgrain structure. Individual dislocations are visible and labelled within Grain~2 (SI Fig.~S2). Counting these yields an estimated dislocation density of $\sim 8 \times 10^{10}~\text{m}^{-2}$. Alternatively, the dislocation density can be estimated from the lattice curvature, using the measured angular spread of $\sim 0.02^{\circ}$ for the $\langle 110 \rangle$ reflection in Grain~2 together with the Burgers vector of $0.248~\text{nm}$ for $\tfrac{1}{2}\langle 111 \rangle$ dislocations on $\{110\}$ planes in bcc Fe, which gives a comparable value. Similar observations have previously been reported in long-annealed aluminum single crystals using DFXM \cite{extensive}.

To further interpret these observations, we compared the local orientation maps with the corresponding axial strain maps for each grain (Fig.~\ref{fig:projection_imaging}C). Although previous studies suggest a potential link between local misorientation and elastic strain fields \cite{DESPRES2024114010}, no correlation was found in our data. To complement the 2D intragranular deformation maps presented in Fig. 2B-C, we imaged Grain 1 and Grain 2 in 3D.


\subsection*{3D DFXM Grain Maping of Selected Grains}

In this section, we demonstrate that the developed method enables 3D grain mapping of embedded and targeted grains. After successfully determining grain orientations via 3DXRD and DCT, we performed 3D DFXM scans on Grain 1 and Grain 2 using two distinct approaches: a layer-by-layer section scan using a incident line beam for grain 2 and MTT \cite{isern2025} scan for Grain 1. Grain 3 was not mapped in 3D.  

\begin{figure*}[h!]
    \centering
    \includegraphics[width=0.6\linewidth]{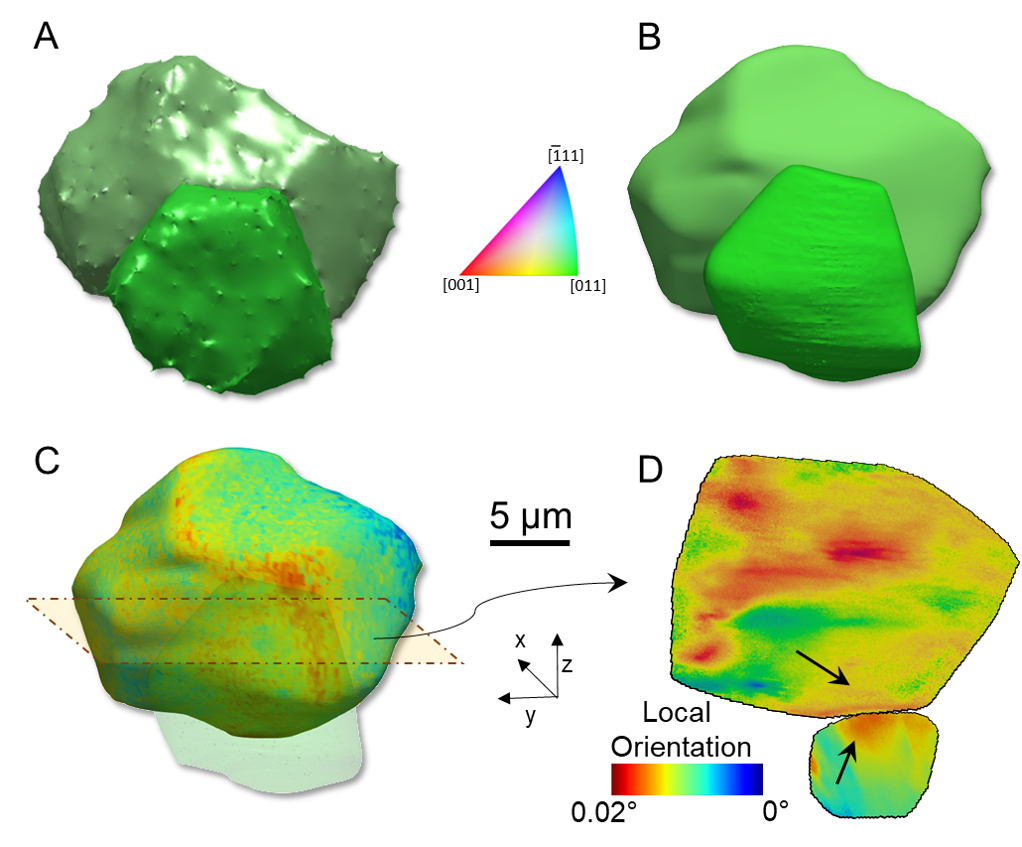}
    \caption{\justifying
    \textbf{Multimodal 3D mapping of neighboring grains using DCT and DFXM.}
    \small 
    (A) DCT-based segmentation of grains 1 and 2. 
    (B) 3D DFXM grain map : G2 (pale green, on the back) imaged via DFXM layer scanning modality, and G1 (bright green, on the front) via magnified topotomography modality of DFXM. 
    (C) Reconstructed local orientation map near the grain boundary interface for G1.
    (D) 2D crossection along the X-Y plane highlighted in (C) showing DFXM local orientation map obtained from layer scanning modality. 
    }
    \label{fig:multiscale}
\end{figure*}

Fig.~\ref{fig:multiscale}A shows the DCT reconstruction of Grain 1 (bright green, front) and Grain 2 (pale green, back).  
Fig.~\ref{fig:multiscale}B shows the corresponding grain shape reconstructions from (i) MTT and (ii) layer-by-layer DFXM scans with the same colorscale in A. Comparing , the grain shapes in Fig. ~\ref{fig:multiscale}A and Fig. ~\ref{fig:multiscale}B, we find that the DCT, MTT and layer-by -ayer reconstruction, all exhibit similar faceted morphologies. This demonstrates the accuracy of the proposed multiscale approach and the reliability of MTT and layer-by-layer grain shape reconstruction modalities. While the effective voxel size for DCT maps is $0.65~\mu\mathrm{m}$, the DFXM grain map has a much smaller pixel size of about $36~\mathrm{nm}$, providing a significant \textit{zoom-in} on the grain structure. The alignment for the grain 2 for MTT measurement was made possible by the simplified alignment procedure outlined in the SI. Compared to the layer-by-layer scan mode, MTT provides improved temporal resolution and higher spatial resolution with isotropic voxel dimensions in the reconstructed image. While MTT allows visualization of grain morphology via tomographic reconstruction, resolving intragranular misorientation remains a challenge and is beyond the scope of this work. To address this limitation here, Grain 1 was additionally scanned using the layer-by-layer mode of DFXM, enabling the acquisition of intragranular misorientation maps. 

Fig.~\ref{fig:multiscale}C shows the 3D intragranular misorientation map of Grain 2 near its boundary with Grain 1, where a localized zone of increased misorientation is clearly visible. While this may be partly explained by the high-angle misorientation ($\sim 53^\circ$) between the grains \cite{HARTE2020555}, similar misorientation patterns are also observed along other segments of the same boundary of Grain~2 (see the S-shaped orange/red region in Fig.~\ref{fig:multiscale}C), suggesting additional factors \cite{RATANAPHAN2015346}. Misorientation accumulation near grain boundaries may be attributed to release of local stresses that may be present due to the thermal history of the sample, including prior annealing treatments \cite{zhang2022local}. A video presenting the 3D rendering of the DFXM grain shapes and misorientation map for Grain 1 and 2 is provided in the SI.

Focusing on the Grain 1–Grain 2 interface, both grains exhibit higher local misorientation near the boundary, likely due to mechanical constraints from neighboring grain interactions. Fig.~\ref{fig:multiscale}D shows a 2D cross-section of the DFXM misorientation map in the XY plane, where misorientation build-up is observed on both sides of the boundary. Black arrows highlight these zones. This result suggests that local lattice rotations in one grain can influence the orientation and mechanical state of adjacent grains. These misorientation gradients are below $0.02^\circ$, within the detection capability of DFXM but beyond that of other methods. Such small variations can significantly influence microstructure evolution under external thermal or mechanical stresses \cite{PhysRevMaterials.5.L070401}. The ability to capture such fine angular variations near targeted grain boundaries with a 100 nm spatial resolution demonstrates the  utility of this multiscale grain mapping workflow. 

\subsection*{Interfacing DFXM and LabDCT}

\begin{figure*}[t]
    \centering
    \includegraphics[width=0.8\linewidth]{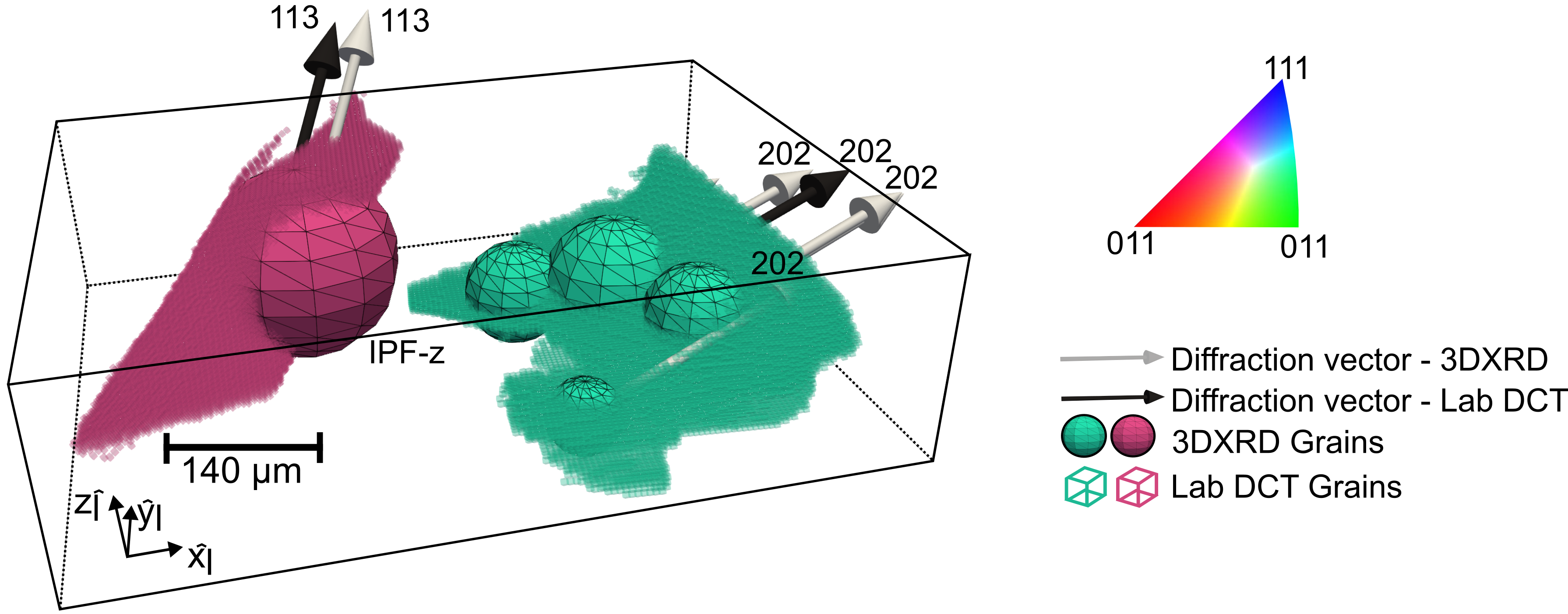}
    \caption{
    \justifying
    \textbf{Comparison of LabDCT and 3DXRD reconstructions of two silicon crystals.} 
    \small The reconstructed 3DXRD grains (shown as spheres) are overlaid onto the LabDCT voxel volumes for visual comparison. 
    The available diffraction vectors for DFXM imaging, as calculated by our alignment algorithm, are indicated as black vectors for LabDCT and white vectors for 3DXRD. 
    }
    \label{fig:labDCTinterfacing}
\end{figure*}

To extend the applicability of our framework beyond synchrotron environments, we validate its interoperability with LabDCT. This capability is essential for bridging pre-characterization at the lab scale with in depth, high-resolution studies at synchrotrons or XFELs. To achieve this, a LabDCT measurement was conducted on a test silicon sample (see Methods for details on samples). The predicted goniometer positions required for a DFXM experiment using the LabDCT measurement were then compared to the predicted goniometer positions from a 3DXRD scan on the same grains.

The reconstructed 3DXRD grains have been overlaid as spherical markers onto the registered LabDCT voxel volumes in Fig.~\ref{fig:labDCTinterfacing} (see Methods for details about the volume registration procedure). Each 3DXRD grain is scaled according to the cube root of its total indexed diffracted intensity to match the equivalent grain radii from the LabDCT data. We used the developed algorithm with accompanying software to interface with both the LabDCT and 3DXRD reconstruction and predict what diffraction vectors would be measurable by DFXM as shown with black (LabDCT) and white (3DXRD) arrows. Note that the green LabDCT grain is identified as 4 different 3DXRD grains, owing to the high angular resolution of 3DXRD. The misorientations between each 3DXRD sub-grain (green spheres) and the LabDCT voxel grain (green volume) are summarized in SI Table 2.

%


The two DFXM predictions were consistent across independent measurements, within the resolution limits reported in Supplementary Table 2. These results show that LabDCT voxel orientation maps, when registered properly, can be used to predict DFXM diffraction peaks with an accuracy of 0.1–0.2$^\circ$. This has important implications as it enables more targeted DFXM experiments and improves experimental throughput by allowing sample pre-screening with lab measurements.


\section*{Conclusion and Outlook}\label{sec:conclusion}

In summary, we present a generalized framework that bridges coarse-grained orientation mapping methods (LabDCT, 3DXRD, DCT, PCT) with high-resolution diffraction imaging via DFXM. Central to this workflow is the open-source software \texttt{crispy}, which converts grain orientations and positions into precise goniometer coordinates, enabling reproducible targeting of thousands of grains within seconds. The framework is robust across both laboratory and synchrotron platforms, and is further extendable to emerging laboratory X-ray techniques such as Lab3DXRD~\cite{oh2025taking} and micro-Laue~\cite{zhang2025laboratory}. This enables non-destructive studies on thousands of grains while deliberately zooming in on selected grains, triple junctions, and clusters to resolve defects and their interactions within the polycrystalline aggregate. Demonstrated on recrystallized bcc iron, the methodology reveals dislocations, subgrains, and orientation gradients in 3D and, by aligning multiple Bragg reflections of a chosen grain, offers a pathway to full strain tensor reconstruction~\cite{Detlefs2025}. This capability establishes a new route for grain mapping that directly links mesoscale microstructure quantification with nanoscale defect visualization, providing an experimental bridge to materials models. The framework is readily transferable to XFELs and laboratory sources, paving the way for integrated, grain-resolved 3D and 4D characterization.

\section*{Methods}

\subsection*{Materials}

The primary sample used in this study was a fully recrystallized high-purity (99.9\%) iron polycrystal. The sample was prepared by annealing material cut from a 90\% cold-rolled sheet at 700\,\textdegree{}C for 30 minutes, followed by electropolishing. The specimen had a wedge-shaped cross-section over an 8\,mm length, with the region of interest approximately $200 \times 200~\mu$m$^2$ located in the thicker section. The average grain size in this region was confirmed to be around 20\,\textmu m by optical microscopy.

To demonstrate offline interoperability with lab-DCT, a secondary sample consisting of three single-crystal silicon shards was studied. These shards were mechanically broken from a silicon wafer and mounted on a Thorlabs magnetic pin holder. The scanned volume for Lab-DCT was approximately $800 \times 800 \times 800~\mu$m$^3$, while the re-measured 3DXRD region at ESRF ID03 covered $800 \times 800 \times 200~\mu$m$^3$. The crystals had anisotropic shapes with equivalent diameters ranging from 200–400\,\textmu m, as measured by Lab-DCT.

\subsection*{Experimental Setup}\label{sec:exp}

All experiments (except LabDCT) were conducted at beamline ID03 of the European Synchrotron Radiation Facility (ESRF)~\cite{isern2025}, which enables combined use of 3DXRD, DCT, PCT and DFXM on the same sample without dismounting or repositioning. This allows for flexible and efficient multiscale characterization within a unified experimental setup.

\subsubsection*{3DXRD Measurements}

3DXRD measurements were performed at 55.12\,keV using a double-crystal monochromator. Prior to sample mounting, detector geometry, energy calibration, and sample-to-detector distance were refined using a silicon powder standard and a 200\,\textmu m$^3$ single-crystal silicon cube.

The pure iron sample was aligned at the calibrated center of rotation on a high-precision goniometer. A full 360$^\circ$ rotation scan was conducted with 0.1$^\circ$ step size using a FReLoN detector (effective pixel size 47.3\,\textmu m, $2048 \times 2048$ resolution) equipped with a fiber taper. Flat-field and dark-field images were acquired for correction.

Calibrations and diffraction data analysis were carried out using \texttt{ImageD11} software package \url{https://github.com/FABLE-3DXRD/ImageD11} to obtain grain positions, orientations, and elastic strain values. The 3DXRD measurement of the silicon sample was conducted using the same procedure as that applied to the high purity iron sample described above.  More about experimental and analysis procedure for 3DXRD can be found elsewhere. \cite{poulsen2004three,ball_revealing_2024,BALL2024119608}

\subsubsection*{DCT Measurements}

DCT measurements were carried out using monochromatic X-rays at 55.12\,keV. A full 360$^\circ$ rotation scan was conducted with 0.1$^\circ$ step size. The scintillator of the near-field detector (PCO.edge 5.5) was positioned 4.5\,mm downstream of the sample, coupled with a 10$\times$ optical objective, yielding an effective pixel size of 0.65\,\textmu m at the sample. Reconstruction of the three dimensional grain volume and segmentation of individual grains were performed using the graintracking MATLAB package developed at ESRF \url{https://gitlab.esrf.fr/graintracking/dct}. Crystallographic orientations for each grain were computed and expressed in the same laboratory reference frame as the 3DXRD data, enabling direct comparison. More about experimental and analysis procedure for DCT can be found elsewhere. \cite{10.1063/1.3100200,Fang:nb5400}

\subsubsection*{Lab-DCT Measurements}

The Lab-DCT measurement of the silicon sample was done on a ZEISS Xradia 520 Versa X-ray microscope integrated with a Lab-DCT imaging module. The sample was placed at a distance of 14 mm from both the X-ray source (tube) and the detector. The X-ray tube was operated at an electron acceleration voltage of 110 keV and a power of 10 W, leading to a polychormatic beam with energy up to 110 KeV. The detector setup included a CDD camera and a scintillator. A total of 2032 x 2032 pixels coupled with a 4x objective lens yielded an effective pixel size of 3.36 $\mu$m. During the Lab-DCT scan, 121 projections were acquired with $3^\circ$ angular steps and 30 s exposure time. In addition to this, an absorption scan was taken to provide the shape of the sample for the reconstruction. Data was analysed using the GrainMapper3D software. Reconstructed 3DXRD and LabDCT volumes were registered by determining the misorientation of the pink grain in Fig.~\ref{fig:labDCTinterfacing} and rotating the LabDCT volume accordingly. More about experimental and analysis procedure for LabDCT can be found elsewhere. \cite{10.1063/1.3100200,Fang:iv5008,Lindkvist:nb5277}

\subsubsection*{DFXM Measurements}

DFXM measurements were performed using monochromatic X-rays of 17~keV. More details about coordinate systems at different measurement modes at ID03 are covered in detail here \cite{isern2025}. A near-field alignment camera positioned 100~mm behind the sample was used to coarsely orient the grain into the Bragg condition. Following the alignment, the camera was removed and imaging proceeded in the far field using a compound refractive lens (CRL) objective.

The objective CRL consisted of 87 beryllium bi-paraboloid lenslets with a radius of curvature $R = 50~\mu$m. The lens was placed approximately 260~mm downstream of the sample, giving a numerical aperture of 0.693~mrad and a magnification of $M_x = 18.3$ at 17~keV, calculated following Equation 9 in Ref.~\cite{Poulsen2017}. The X-ray image formed by the CRL was projected onto a far-field detector, which employed an indirect detection scheme consisting of a scintillator, visible-light optics, and a PCO.edge 4.2 bi sCMOS camera ($2048 \times 2048$ pixels). The detector was positioned 5348~mm from the sample. Optical magnification of  $10\times$ resulted in an  effective pixel size of 36.3~nm. The 110 reflection was selected for all DFXM scans.

Three types of DFXM scans were performed: (i) projection scans using a parallel beam shaped to $300 \times 300~\mu$m$^2$ via slits, (ii) section scans using a line-focused beam generated by a 1D Be- condenser CRL comprising 58 lenses with a radius of curvature $R = 100~\mu$m positioned 718 mm upstream of the sample (resulting in a $200 \times 0.6~\mu$m$^2$ beam profile), and (iii) magnified topotomographic scans (MTT) using a box-beam.

Rocking scans were performed by scanning the $\mu$ axis over narrow angular intervals, acquiring images that reveal local orientation variations. Strain scans were collected by varying the $\phi$ and $2\theta$ angles within ranges of $0.1^\circ$ and $0.08^\circ$, respectively, while keeping $\chi$ fixed. In projection mode, entire grains were imaged at a fixed $\omega$ position using full-field magnification, with simultaneous acquisition of rocking and strain scans. In section mode, virtual two-dimensional slices of selected grains were obtained by vertically translating the sample through the narrow focal plane while performing only the rocking scan. For the MTT mode, the diffraction vector was aligned parallel to the rotation axis using the procedure outlined in supplementary information, enabling the acquisition of diffraction projections across 360$^\circ$ without the need to reconstruct from 2D layer sections. 720 angular projections were acquired to reconstruct the grain shape. More on data analysis can be found here \cite{darfix}. 

A total of three neighboring grains were examined using DFXM. All were measured in projection modes to obtain DFXM orientation and strain maps. Additionally, G1 was measured in section and MTT mode. G2 was measured in section mode. All DFXM data was reconstructed using open source python package \texttt{Darling} \url{https://github.com/AxelHenningsson/darling}. 

\subsection*{Grain Alignment for DFXM}

A subset of grains located near the volume center of the pure iron sample was selected for DFXM imaging. Using the open source \texttt{crispy} package, goniometer angles were calculated to bring each grain into the Bragg condition for the 110 reflection at 17\,keV. The goniometer motor bounds were set to the following ranges: $\mu$ from $-3^\circ$ to $15^\circ$, $\omega$ from $-179^\circ$ to $179^\circ$, and both $\chi$ and $\phi$ from $-6^\circ$ to $6^\circ$. The detector position was constrained along the $z$-axis between $-0.04\,\mathrm{m}$ and $1.96\,\mathrm{m}$. The predicted DFXM positions at $\eta = 0$ showed excellent agreement with experimental values,within 0.05$^\circ$ in $\mu$ and 0.01$^\circ$ in both $\chi$ and $\phi$, demonstrating the reproducibility and precision of the alignment workflow, particularly in the low-mosaicity recrystallized microstructure.

To demonstrate interoperability of the software between 3DXRD results and LabDCT results, the predicted goniometer positions required for a DFXM experiment using the LabDCT measurement on the silicon sample were then compared to the predicted goniometer positions from a 3DXRD scan on the same grains. The theoretical goniometer motor bounds were set to the following ranges: $\mu$ from $0^\circ$ to $15^\circ$, $\omega$ from $-179^\circ$ to $179^\circ$, and both $\chi$ and $\phi$ from $-7.5^\circ$ to $7.5^\circ$. The detector position was constrained along the $z$-axis between $-0.04\,\mathrm{m}$ and $1.96\,\mathrm{m}$. We used a detector-to-sample distance of $4\,\mathrm{m}$ and an X-ray energy of $17\,\mathrm{keV}$. As annotated in Fig.~\ref{fig:labDCTinterfacing} the 113 and 202 reflections were found to be accessible in this experimental setup.

\section*{Acknowledgement}
We thank ESRF for providing the beamtime at ID03. CY and AS acknowledge the financial support from the ERC Starting Grant nr 10116911. We acknowledge the technical help provided by H. Isern and T. Dufrane during the experiments. We thank H.F. Poulsen for discussions to improve the manuscript. AAWC, NAH and MLB acknowledge support from the ERC Advanced Grant nr 885022 and from the Danish ESS lighthouse on hard materials in 3D, SOLID. 

\section*{Data Availability}

The open-source analysis package \texttt{crispy}, developed for calculating goniometer positions from 3DXRD grain orientation data, is freely available at \url{https://github.com/AxelHenningsson/crispy}. Experimental data for this paper is available at https://doi.org/10.15151/ESRF-ES-2128361408. A curated subset of raw and processed 3DXRD and DFXM datasets used in the figures will be made available via a public data archive upon publication.

\bibliography{references}

\end{document}